\UseRawInputEncoding

\documentclass[12pt]{article}
\usepackage{longtable,supertabular}
\usepackage{amsmath}
\usepackage{amssymb}
\usepackage{latexsym}
\usepackage{cite}

\title{\bf Many Non-Reed-Solomon Type MDS Codes From Arbitrary Genus Algebraic Curves}
\author{Hao Chen
  \thanks{Hao Chen is with the College of Information Science and Technology/Cyber Security, Jinan University, Guangzhou, Guangdong Province, 510632, China, haochen@jnu.edu.cn. The research of Hao Chen was supported by NSFC Grant 62032009.}}

\begin{document}

\maketitle
\begin{abstract}
It is always interesting and important to construct non-Reed-Solomon type MDS codes in coding theory and finite geometries. In this paper, we prove that there are non-Reed-Solomon type MDS codes from arbitrary genus algebraic curves.  It is proved that MDS algebraic geometry (AG) codes from higher genus curves are not equivalent to MDS AG codes from lower genus curves. For genus one case, we construct MDS AG codes of small consecutive lengths from elliptic curves. New self-dual MDS AG codes over ${\bf F}_{{2^s}}$ from elliptic curves are also constructed. These MDS AG codes are not equivalent to Reed-Solomon codes, not equivalent to known MDS twisted Reed-Solomon codes and not equivalent to Roth-Lempel MDS codes.\\

Hence many non-equivalent MDS AG codes, which are not equivalent to Reed-Solomon codes and known MDS twisted-Reed-Solomon codes, can be obtained from arbitrary genus algebraic curves. It is interesting open problem to construct explicit longer MDS AG codes from maximal curves.\\

\end{abstract}

\section{Introduction and Preliminaries}

The Hamming weight $wt({\bf a})$ of a vector ${\bf a} \in {\bf F}_q^n$ is the number of non-zero coordinate positions. The Hamming distance $d({\bf a}, {\bf b})$ between two vectors ${\bf a}$ and ${\bf b}$ is defined to be the Hamming weight of $wt({\bf a}-{\bf b})$. For a code ${\bf C} \subset {\bf F}_q^n$, its minimum Hamming distance $$d({\bf C})=\min_{{\bf a} \neq {\bf b}} \{d({\bf a}, {\bf b}),  {\bf a} \in {\bf C}, {\bf b} \in {\bf C} \}$$  is the minimum of Hamming distances $d({\bf a}, {\bf b})$ between any two different codewords ${\bf a}$ and ${\bf b}$ in ${\bf C}$. For a linear code ${\bf C} \subset {\bf F}_q^n$, its minimum Hamming distance is its minimum Hamming weight. For a linear $[n, k, d]_q$ code, the Singleton bound asserts $d \leq n-k+1$. When equality holds, this code is an MDS code. We refer to \cite{HP,MScode} for the theory of error-correcting codes. In \cite[Page 317]{MScode}, the Chapter 11 "MDS codes"  is called "one of the most fascinating chapters in all of coding theory".\\

The main conjecture of linear MDS codes proposed in \cite{Segre} claims that the length of a linear MDS code over ${\bf F}_q$ is at most $q+1$, except some exceptional cases. In \cite{Ball} the main conjecture was proved for linear MDS codes over prime fields. Some classification results about general MDS codes over small fields were given in \cite{KKO}. An $(n, M=q^{n-d}, d)_q$ code is called almost MDS.  A linear almost MDS code ${\bf C}$ satisfying that the dual ${\bf C}^{\perp}$ is also almost MDS is called near MDS code. It is well-known that AG codes from elliptic curves are near MDS codes. The main conjecture of near MDS codes was proposed in \cite{Landjev}. For counting the number of MDS linear codes, we refer to \cite{Ghorpade,Kaipa}.\\

We say that two codes ${\bf C}_1$ and ${\bf C}_2$ in ${\bf F}_q^n$ are equivalent if ${\bf C}_2$ can be obtained from ${\bf C}_1$ by a permutation of coordinates and the multiplication of a Hamming weight $n$ vector ${\bf v}=(v_1, v_2, \ldots, v_n) \in {\bf F}_q^n$ on coordinates, where $v_i \neq 0$ for $i=1, \ldots, n$. That is $${\bf C}_2=\{{\bf c}=(c_1, \ldots c_n): (c_1, \ldots, c_n)=(v_1x_1, \ldots, v_nx_n), {\bf x} \in Perm({\bf C}_1)\},$$ where $Perm({\bf C}_1)$ is the code obtained from ${\bf C}_1$ by a coordinate permutation. Equivalent codes have the same code length, the dimension and the minimum Hamming distance.\\

Reed-Solomon codes proposed in the 1960 paper \cite{RS} are well-known MDS codes. Let $P_1,\ldots,P_n$ be $n \leq q$ distinct elements in ${\bf F}_q$. The Reed-Solomon code $RS(n, k)$ is defined by $$RS(n,k)=\{(f(P_1),\ldots,f(P_n)): f \in {\bf F}_q[x],\deg(f) \leq k-1\}.$$ This is a $[n,k,n-k+1]_q$ linear MDS codes, because a degree $\deg(f) \leq k-1$ polynomial has at most $k-1$ roots. These codes were called "the greatest codes of them all" in \cite[Chapter 5]{GRS}. Reed-Solomon codes are AG codes from the genus zero curve. It should be mentioned that Reed-Solomon codes were widely used in secret sharing, secure multiparty computation, see \cite{Shamir,McEliece,BGW,CCD,CC06} and distributed storage systems, see \cite{CJM}.\\

MDS codes which are not equivalent to Reed-Solomon codes are called non-Reed-Solomon type MDS codes. It is always interesting and important to construct non-Reed-Solomon MDS codes in coding theory and finite geometries, we refer to \cite{RL1989,BGP,BPN17,BPR}. One method used to distinguish MDS codes from Reed-Solomon codes is the calculation of dimensions of their Schur squares, see \cite{BPN17,BPR}. The componentwise product (star product) of $t$ vectors ${\bf x}_j=(x_{j,1}, \ldots, x_{j, n}) \in {\bf F}_q^n$, $j=1, \ldots, t$, is ${\bf x}_1 \star \cdots \star {\bf x}_t=(x_{1,1} \cdots x_{t,1}, \ldots, x_{1, n} \cdots x_{t,n}) \in {\bf F}_q^n$. The componentwise product of linear codes ${\bf C}_1, \ldots, {\bf C}_t$ in ${\bf F}_q^n$ is defined by $${\bf C}_1\star \cdots \star {\bf C}_t=\Sigma_{ {\bf c}_i \in {\bf C}_i} {\bf F}_q {\bf c}_1 \star \cdots \star {\bf c}_t. $$ When ${\bf C}_1={\bf C}_2={\bf C}$, the componentwise product ${\bf C} \star {\bf C}$ is called the Schur square of the linear code ${\bf C}$. It is clear that Schur squares of equivalent linear codes are equivalent. Hence the dimension and the minimum Hamming distance of Schur squares are invariants of two equivalent linear codes. The dimension of the Schur square of the Reed-Solomon $[n, k, n-k+1]_q$, $2k \leq n \leq q+1$, is $2k-1$. Then any MDS $[n, k, n-k+1]_q$ code ${\bf C}$ satisfying $\dim({\bf C} \star {\bf C}) \geq 2k$, is a non-Reed-Solomon type MDS code, see \cite{BPR}.\\

The interest to construct non-Reed-Solomon type MDS codes comes from the invention of the twisted Gabidulin codes in the rank-metric. Rank-metric codes attaining the similar Singleton bound $|{\bf C}| \leq q^{n(n-d_r({\bf C})+1)}$ is called maximal rank distance (MRD) codes, where $d_r({\bf C})$ is the minimum rank-distance of the rank-metric code ${\bf C}$. Gabidulin codes are well-known MRD codes and can be thought as the rank-metric analogue of Reed-Solomon codes, see \cite{Gabidulin}. It is a great achievement that some twisted Gabidulin codes were found in \cite{Sheekey}. These rank-metric codes are MRD codes, which are not equivalent to Gabidulin codes, see \cite{Sheekey,Neri}. Then many papers, see  e.g., \cite{CMPZ,LTZ,TZ}, were published for constructing new twisted Gabidulin codes. \\

Motivated by the progress of twisted Gabidulin codes, twisted Reed-Solomon codes were introduced by P. Beelen, S. Puchsinger, J. Nielson and J. Rosenkilde in \cite{BPN17,BPR}. Some of these twisted Reed-Solomon codes are MDS code, see Section 4 of \cite{BPR} and \cite{LL21}, and MDS twisted Reed-Solomon codes are non Reed-Solomon type MDS codes. However it seems that there are strong restrictions on lengths of these MDS twisted Reed-Solomon codes, see \cite{BPR}. In fact, in an earlier paper \cite{RL1989}, some non-Reed-Solomon type MDS codes were constructed and studied. Twisted Hermitian codes from Hermitian curves were introduced in a recent paper \cite{ABFKMMN}.\\

The (Euclid) dual of a linear code ${\bf C}\subset {\bf F}_q^n$ is ${\bf C}^{\perp}=\{{\bf c}=(c_1, \ldots, c_n): \Sigma_{i=1}^n c_i x_i=0, \forall {\bf x}=(x_1, \ldots, x_n) \in {\bf C}\}$. A linear code is called self dual if ${\bf C}={\bf C}^{\perp}$. In general the linear code ${\bf C} \bigcap {\bf C}^{\perp}$ is called the hull of ${\bf C}$. The Hermitian dual of a linear code ${\bf C} \subset {\bf F}_{q^2}^n$ is $${\bf C}^{\perp_h}=\{{\bf c}=(c_1, \ldots, c_n): \Sigma_{i=1}^n c_i x_i^q=0, \forall {\bf x}=(x_1, \ldots, x_n) \in {\bf C}\}.$$  It is clear ${\bf C}^{\perp_h}=({\bf C}^{\perp})^q$, where $${\bf C}^q=\{(c_1^q, \ldots, c_n^q): (c_1 \ldots, c_n) \in {\bf C}\}.$$ The minimum distance of the Euclid dual is called the dual distance and is denoted by $d^{\perp}$. The minimum distance of the Hermitian dual is the same as $d^{\perp}$. A linear code ${\bf C} \subset {\bf F}_{q^2}^n$ is called Hermitian self-dual if ${\bf C}={\bf C}^{\perp_h}$. The intersection ${\bf C} \bigcap {\bf C}^{\perp_h}$ is called the Hermitian hull of this code ${\bf C}$. We refer to \cite{CPS,CS} and \cite[Chapter 9]{HP} for earlier results about self-dual and Hermitian self-dual codes over small fields. From the Calderbank-Shor-Steane (CSS) construction of entanglement-assisted quantum error correction (EAQEC) codes in  \cite{Brun}, Euclidean and Hermitian self-dual MDS codes can be used to construct MDS EAQEC codes.\\

The construction of new self-dual MDS codes or near MDS codes has been a long active topic in coding theory, see \cite{Gulliver,Bassa,HY20,JX17,JK19,Niu,ZhangFeng,GuoLi,FangLiuLuo}. On the other hand the construction of Hermitian self-orthogonal (or dual-containing) MDS codes had been active for the purpose to construct MDS quantum codes, see \cite{HCX,KZ,Ball1,Ball2,Ball3} and references therein. Since the introduction of twisted Reed-Solomon codes in \cite{BPN17,BPR}, the construction of non-Reed-Solomon self-dual MDS codes from twisted Reed-Solomon codes has been given in \cite{HY20,SYLH}. These codes are not equivalent to the Reed-Solomon codes and can be thought as new self-dual MDS codes. From the view of coding theory, it is always interesting to construct non-Reed-Solomon type MDS codes and non-Reed-Solomon type self-dual MDS codes.\\

Let ${\bf X}$ be an absolutely irreducible, smooth and genus $g$ curve defined over ${\bf F}_q$. Let ${\bf P}=\{P_1,\ldots,P_n\}$ be the set of $n$ distinct rational points of ${\bf X}$ over ${\bf F}_q$. Let ${\bf G}$ be a rational divisor over ${\bf F}_q$ of degree $\deg({\bf G})$ satisfying $2g-2 <\deg({\bf G})<n$ and $$support({\bf G}) \bigcap {\bf P}=\emptyset.$$ Let ${\bf L}({\bf G})$ be the function space associated with the divisor ${\bf G}$, that is, ${\bf L}({\bf G})$ is the space of all rational functions $f$ satisfying $(f)+{\bf G} \geq 0$, where $(f)$ is the divisor associated with $f$. The algebraic geometry code (functional code) associated with ${\bf G}$, ${\bf P}=\{P_1,\ldots,P_n\}$ is defined by $${\bf C}({\bf P}, {\bf G}, {\bf X})={\bf C}(P_1,\ldots,P_n, {\bf G}, {\bf X})=\{(f(P_1),\ldots,f(P_n)): f \in {\bf L}({\bf G})\}.$$ The dimension of this code is $$k=\deg({\bf G})-g+1$$ follows from the Riemann-Roch Theorem. The minimum Hamming distance is $$d \geq n-\deg({\bf G}).$$ Algebraic-geometric residual code ${\bf C}_{\Omega}(P_1, \ldots, P_n, {\bf G}, {\bf X})$ with the dimension $k=n-m+g-1$ and minimum Hamming distance $d \geq m-2g+2$ can be defined, we refer to \cite{HP,TV} for the detail. It is the dual code of the functional code of the dimension $m-g+1$.  The AG codes from elliptic curves satisfy  $k+d \geq \deg({\bf G})+n-\deg({\bf G})=n$. Hence these elliptic curve codes are near MDS codes.\\

A divisor ${\bf G}=\Sigma m_i G_i$ where $G_i$'s are points of the curve, is called effective if $m_i \geq 0$. Two effective divisor ${\bf G}_1$ and ${\bf G}_2$ are called linear equivalent if there is a rational function $f$ such that the divisor $(f)$ associated with $f$ is of the form $$(f)={\bf G}_1-{\bf G}_2.$$ It is clear that for two linear equivalent divisors ${\bf G}_1$ and ${\bf G}_2$, the AG codes ${\bf C}(P_1, \ldots, P_n, {\bf G}_1, {\bf X})$ and ${\bf C}(P_1, \ldots, P_n, {\bf G}_2, {\bf X})$ are equivalent linear codes.\\

It is well-known that AG codes from elliptic curves are best examples of linear codes with the Singleton defect $n+1-d-k=1$. However it is also well-known that MDS codes can be obtained from elliptic curve codes if the evaluation points are carefully chosen. This is similar to many works on Reed-Solomon codes in which evaluation points have to be determined in a complicated pattern, see \cite{GRS} and references therein. The main conjecture for MDS codes from curves of genus $1$ (elliptic) and $2$ was proved in old papers \cite{Munuera1,Munuera}, also see \cite{Walker}. The main conjecture for MDS AG codes arising from hyper-elliptic curves was proved independently in \cite{Chen2} and \cite{Boer}. For a tighter upper bound on the maximal length of MDS codes from elliptic curves, we refer to a recent paper \cite{HR}.  The self-dual near MDS codes from elliptic curves were constructed in \cite{JK19} and some non-extendable near MDS codes from elliptic curves were constructed in a recent paper \cite{AGS}. We also refer to \cite{DL} for the earlier work on near MDS codes. In \cite{BGP} some MDS $[n, k, n-k+1]_q$ codes with the covering radius $n-k-1$ were constructed as extended codes of MDS codes from elliptic curves. It is well-known that the covering radius of the Reed-Solomon $[n, k, n-k+1]_q$ code is $n-k$. Therefore these MDS codes in \cite{BGP} are non-Reed-Solomon type MDS codes. It is interesting and important to ask the following question about MDS AG codes.\\

{\bf Problem.} {\em Can MDS AG codes be obtained from arbitrary genus curves by choosing evaluation points carefully? Are these MDS AG codes non-Reed-Solomon type? Moreover are these MDS AG codes equivalent to each other?}\\

In this paper, we prove that there are many non-Reed-Solomon type MDS codes over ${\bf F}_q$  from arbitrary genus curves. It is proved that MDS AG codes from genus $g_1$ curves are not equivalent to MDS AG codes from genus $g_2$ curves, if $g_1>g_2$. In the range of lengths $n$ satisfying $n \leq q^{1/4}$ and $\gcd(n, q)=1$, non-Reed-Solomon type MDS elliptic curve codes can be constructed for consecutive lengths. These codes can be constructed as AG codes from elliptic curves ${\bf E}$ defined over ${\bf F}_q$ by choosing the evaluation point set as a coset of a subgroup of  ${\bf E}({\bf F}_q)$. They are not equivalent to Reed-Solomon codes, not equivalent to the known MDS twisted Reed-Solomon codes and not equivalent to Roth-Lempel MDS codes in \cite{RL1989}. Moreover there are non-Reed-Solomon type MDS elliptic curve codes with the same length and the same dimension. Many new self-dual non-Reed-Solomon type MDS codes with various lengths are also constructed. It is always good to understand that algebraic geometry codes can give us more interesting non-Reed-Solomon type MDS codes.\\

\section{AG codes from elliptic curves}

\subsection{Elliptic curves over finite fields}

In this subsection, we recall basic facts about elliptic curves defined over a finite field, which are mainly from the paper \cite{Ruck}. \\

Let ${\bf E}$ be an elliptic curve defined over ${\bf F}_q$.  It is well-known that when $q$ is not a power of $2$ or $3$, then elliptic curves over ${\bf F}_q$ can be realized as a non-singular plane cubic curve. Let ${\bf E}({\bf F}_q)$ be the set of all ${\bf F}_q$-rational points of ${\bf E}$. The number $|{\bf E}({\bf F}_q)|$ of its rational points over ${\bf F}_q$ satisfies the Hasse bound $$|q+1-|{\bf E}({\bf F}_q)|| \leq 2\sqrt{q}.$$  For any positive real number $x$ we set $$x^{-}=x+1-2\sqrt{x}, $$ and $$x^{+}=x=1+2\sqrt{x}.$$ If $q=p$ is a prime number it follows from the result in \cite{Deuring,Ruck} that for any positive integer $N$ satisfying $p^{-} <N < p^{+}$, there is an elliptic curve ${\bf E}$ defined over ${F}_p$ such that the number of ${\bf F}_p$-rational points of ${\bf E}$ satisfying $$|{\bf E}({\bf F}_p)|=N.$$

It is well-known there is an Abelian group structure on  ${\bf E}({\bf F}_q)$. As a group ${\bf E}({\bf F}_q) \simeq {\bf Z}/m{\bf Z} \times {\bf Z}/mk{\bf Z}$ for some positive integers $m$ and $k$. We refer to \cite{Silverman,Ruck} for the detail. More accurately the following two results were proved in \cite{Ruck}.\\

{\bf Theorem 2.1 (Theorem 1a in \cite{Ruck}).} {\em  All the possible orders $|{\bf E}({\bf F}_q)|$ of an elliptic curve ${\bf E}$ defined over ${\bf F}_q$, where $q=p^n$ is a prime power,   are given by $$ |{\bf E}({\bf F}_q)|=1+q-\beta,$$
where $\beta$  is an integer with  $|\beta| \leq 2\sqrt{q}$  satisfying one of the following conditions:\\
(a) $\gcd(\beta,p)=1$;\\
(b) If $n$ is even: $\beta=\pm2\sqrt{q}$;\\
(c) If $n$ is even and $p \neq 1$ $mod$ $3$: $\beta=\pm\sqrt{q}$;\\
(d) If $n$ is odd and $p=2$ or $3$: $\beta=\pm p^{\frac{n+1}{2}}$;\\
(e) If either (i) $n$ is odd or (ii) $n$ is even, and $p\neq 1$ $mod$ $4$: $\beta=0$.}\\

All possible group structures of elliptic curves over ${\bf F}_q$ were also determined in \cite{Ruck}.\\

{\bf Theorem 2.2 (Theorem 3 in \cite{Ruck}).} {\em Let ${\bf E}$ be an elliptic curve over a finite field ${\bf F}_q$  with $q=p^n$  elements.  Let $|{\bf E}({\bf F}_q)|= \prod_{l} l^{h_l}$ be the prime factoring.  Then all the possible groups ${\bf E}({\bf F}_q)$ with the order $|{\bf E}({\bf F}_q)|$ are the following, $${\bf Z}/p^{h_p}{\bf Z} \times \prod_{l\neq p} ({\bf Z}/l^{a_l}{\bf Z}\times {\bf Z}/l^{h_l-a_l}{\bf Z}),$$
with\\
(a) In case (b) of Theorem 3.1: Each $a_l$  is equal to $\frac{h_l}{2}$;\\
(b) In cases (a), (c), (d), (e) of Theorem 3.1: $a_l$  is an arbitrary integer satisfying
$$0 \leq a_l \leq \min \{v_l(q-1), [\frac{h_l}{2}]\},$$ where $v_l(q-1)$ is the order of prime factor $l$ in $q-1$.}\\

From Theorem 2.1 and 2.2, there are a lot of elliptic curves with different orders $|{\bf E}({\bf F}_q)|$ and different Abelian group structures. Therefore we can construct many MDS elliptic curve codes from the following Theorem 2.3.\\

\subsection{MDS codes from elliptic curves}

The following basic facts about AG codes from elliptic curves are well-known, for example, see \cite{Munuera,BGP}. Let ${\bf E}$ be an elliptic curve defined over the finite field ${\bf F}_q$. Let $P_0, P_1, \ldots, P_n$ be $n+1$ rational points. We take the divisor ${\bf G}=mP_0$, where $0<m \leq n-1$. Then an AG code ${\bf C}(P_1, \ldots, P_n, {\bf G}, {\bf E})$ from this elliptic curve ${\bf E}$ is constructed. This is a linear $[n, m, \geq n-m]_q$ code.\\

{\bf Theorem 2.3.} {\em 1) The minimum Hamming distance of  ${\bf C}(P_1, \ldots, P_n, {\bf G}, {\bf E})$ is $n-m$ if and only if there are $m$ distinct points $P_{i_1}, \ldots, P_{i_m}$ in the set $\{P_1, \ldots, P_n\}$ such that the effective divisor $P_{i_1}+\cdots+P_{i_m}$ is linear equivalent to the divisor $mP_0$.\\
2) If there is an effective rational divisor ${\bf G}_1$ satisfying that ${\bf G}_1+{\bf G}$ is linearly equivalent to the effective divisor $P_1+\cdots+P_n$. The the dual code of ${\bf C}(P_1, \ldots, P_n, {\bf G}, {\bf E})$ is equivalent to the AG code ${\bf C}(P_1, \ldots, P_n, {\bf G}_1, {\bf E})$.}\\

{\bf Proof.} 1) is direct from the definition of AG code. 2) is direct from the definition of residual code, see \cite{Munuera}.\\

Notice that the Jacobian of the elliptic curve ${\bf E}$ is itself, so the condition in Theorem 2.3 1) is equivalent to if $P_{i_1}+\cdots+P_{i_m}=mP_0$ is valid in the group ${\bf E}({\bf F}_q)$, we refer to \cite{Munuera}.\\

\subsection{The Schur squares of elliptic curve codes}

Let ${\bf E}$ be an elliptic curve defined over ${\bf F}_q$ and $P_0$ is a rational point. We now discuss some properties of the linear space $L(mP_0)$ of rational functions $f$ satisfying $(f)+mP_0 \geq 0$. The following property is well-known.\\

{\bf Proposition 2.1.} {\em ${\bf L}(P_0)$ is the one dimension linear space of all constant functions. For each $m \geq 2$, there is one rational function $f$ in ${\bf L}(mP_0)/{\bf L}((m-1)P_0)$.}\\

{\bf Proof.} If there is a non-constant rational function $f$ in the space ${\bf L}(P_0)$, then $(f)=Q-P$ for some rational point $Q \in {\bf E}({\bf F}_q)$. That is, $P_0$ is linearly equivalent to another rational point $Q \in {\bf E}({\bf F}_q)$. This is a contradiction to the fact that ${\bf E}({\bf F}_q)$ is the Jacobian $$\{{\bf G}: \deg({\bf G}=0\}/linear-equivalence$$ of ${\bf E}$.\\

On the other hand from the Riemann-Roch theorem, $\dim({\bf L}(mP_0))=m$ when $m \geq 2g-1=1$. The second conclusion follows immediately.\\

{\bf Theorem 2.4.} {\em The dimension of the Schur square of an one-point AG code ${\bf C}(P_1, \ldots, P_n, mP_0, {\bf E})$ from an elliptic curve ${\bf E}$ is $2m$ if $6 \leq 2m \leq n$.}\\

{\bf Proof.} Let $f_1$ be the constant function in ${\bf L}(mP_0)$, $f_i$, $i=2, \ldots, m$ be the rational function with the $i$-th order pole at the point $P_0$. It is clear that ${\bf C}(P_1, \ldots, P_n, mP_0, {\bf E})=\{(f(P_1), \ldots, f(P_n)): f= \Sigma a_if_I, a_i \in {\bf F}_q\}$. It is clear that the Schur square of ${\bf C}(P_1, \ldots, P_n, mP_0, {\bf E})$ is in the AG code ${\bf C}(P_1, \ldots, P_n, 2mP_0, {\bf E})$. On the other hand it is easy to verify that the evaluation codewords of  $f_1, f_2, \ldots, f_m$ are in the Schur square. The rational function $f_i f_j$, where $i \geq 2$, $j\geq 2$ satisfying $m+1 \leq i+j \leq 2m$, can have $w$-th order pole at the point $P_0$ for $w=m+1, \ldots, 2m$. Then the Schur square of ${\bf C}(P_1, \ldots, P_n, mP_0, {\bf E})$ is ${\bf C}(P_1, \ldots, P_n, 2mP_0, {\bf E})$. The conclusion is proved.\\

It is obvious that the minimum Hamming distance of the Schur squares of an elliptic curve $[n, k, \geq n-k]_q$ code is at least $n-2k$. Theorem 2.4 can be generalized to one point AG codes from higher genus curves, see Theorem 4.1.\\

\section{Non-Reed-Solomon type MDS codes from elliptic curves}

Let $q$ be a prime power. In this section, the set ${\bf P}=\{P_1, \ldots, P_n\}$ is the disjoint union of several cosets of a subgroup of ${\bf E}({\bf F}_q)$. We show that there are many MDS AG codes of consecutive lengths from elliptic curves ${\bf E}$ defined over ${\bf F}_q$. These MDS codes are not equivalent to the Reed-Solomon codes from dimensions of their Schur squares. On the other hand, MDS twisted Reed-Solomon codes have been only constructed for some special lengths $n$ satisfying that $n$ is a divisor of $q-1$ or $\gcd(n, q-1)=\frac{n}{2}$, or $n$ is a divisor of $q$, see \cite{BPR}. Therefore it is obvious that there are more non-Reed-Solomon MDS AG codes from elliptic curves than twisted Reed-Solomon codes. In the following part, $P_0$ is the zero element in the group ${\bf E}({\bf F}_q)$.\\

{\bf Theorem 3.1.} {\em 1) Let ${\bf E}_1 \subset {\bf E}({\bf F}_q)$ be a subgroup of the order $n_1$ and $b \in {\bf E}({\bf F}_q)$ be a nonzero element such that the $n_2$ order cyclic subgroup $<b>$  generated by $b$ intersects ${\bf E}_1$ at the zero element. Then for any positive integer $m \leq n_2-1$, the elliptic curve code ${\bf C}(b+{\bf E}_1, mP_0, {\bf E})$ is an MDS code.\\
2) Let ${\bf E}_1 \subset {\bf E}({\bf F}_q)$ be a subgroup of the order $n_1$ and $b_i \in {\bf E}({\bf F}_q)$, $i=1, \ldots, t$,  be $t$ nonzero elements in ${\bf E}({\bf F}_q)$ such that the set $\{m_1b_1+\cdots+m_tb_t: m_1+\cdots+m_t=m\}$ intersects the subgroup ${\bf E}_1$ at the zero element. Set ${\bf P}$ be the union of $t$ cosets $b_1+{\bf E}_1, \ldots, b_t+{\bf E}_1$. Then the elliptic curve code ${\bf C}({\bf P}, mP_0, {\bf E})$ is an MDS code.}\\

{\bf Proof.} The sum $P_{i_1}+\cdots+P_{i_m}$ of $m$ rational points in the coset $b+{\bf E}_1$ is of the form $mb+P$, where $P \in {\bf E}_1$. From the condition that the the order $n_2$ cyclic subgroup $<b> \subset {\bf E}({\bf F}_q)$ intersects ${\bf E}_1$ only at zero element, this sum is not zero element in ${\bf E}({\bf F}_q)$, since $m \leq n_2-1$. The conclusion follows directly. The second conclusion can be proved similarly.\\

{\bf Example 3.1.} Let $q=p$ be a prime number, from the classical result in \cite{Deuring} for any given positive integer satisfying $p^{-} \leq N \leq p^{+}$, there is an elliptic curve ${\bf E}$ defined over ${\bf F}_p$ such that $N=|{\bf E}({\bf F}_p)|$. Suppose that $N=p_1p_2$ be the product of two different prime numbers, where $\gcd(p, p_i)=1$ for $i=1, 2$. There is an elliptic curve defined over ${\bf F}_p$ such that the group structure of ${\bf E}({\bf F}_p)$ is ${\bf Z}/p_1 {\bf Z} \oplus {\bf Z}/p_2{\bf Z}$ from Theorem 2.2. Let ${\bf E}_1$ be the cyclic subgroup ${\bf Z}/p_1{\bf Z} \times {\bf 0}$ of the order $p_1$. Then for each element of the form $({\bf 0}, x) \in {\bf 0} \times {\bf Z}/p_2{\bf Z}$, the elliptic curve code ${\bf C}(b+{\bf E}_1, mP_0, {\bf E})$ is an MDS code for $1\leq m \leq p_1$. \\

Let $b_1< \cdots < b_t$ be $t (\leq p_2-1)$ distinct nonzero elements in ${\bf 0} \times {\bf Z}/p_2{\bf Z}$ and $m$ be a positive integer satisfying $mb_t \leq p_2$. Then $m_1b_1+\cdots+m_tb_t\leq mb_t <p_2$, where $m_1+\cdots+m_t=m$. Let ${\bf P}$ be the union of $t$ cosets $b_1+{\bf E}_1, \ldots, b_t+{\bf E}_1$. The elliptic curve code ${\bf C}({\bf P}, mP_0, {\bf E})$ is an MDS code.\\

It is obvious when $t\geq \frac{p_2+1}{2}$, then $m$ cannot bigger than or equal to two. Hence if we want to construct a dimension $2$ MDS elliptic curve code in this example, the length is smaller than $\frac{p_1p_2}{2} \leq q+1$.\\

On the other hand we can take ${\bf E}_1$ be the subgroup ${\bf 0} \times {\bf Z}/p_2{\bf Z}$, then for ${\bf 1} \times {\bf 0}=b$, the elliptic curve code ${\bf C}({\bf P}, mP_0, {\bf E})$ is an MDS code when $m \leq p_1-1$. \\

When $p=19$, we can get the following non-Reed-Solomon type MDS codes, an MDS $[4, 2, 3]_{19}$ code from an elliptic curve with $12$ rational points, an MDS $[5, 2, 4]_{19}$ code from an elliptic curve with $15$ rational points, an MDS $[6, 2, 5]_{19}$ code from an elliptic curve with $18$ rational points, an MDS $[5, 2, 4]_{19}$ code from an elliptic curve with $20$ rational points, and an MDS $[6, 3, 4]_{19}$ code from an elliptic curve with $24$ rational points.\\

{\bf Corollary 3.1.} {\em Let ${\bf E}$ be an elliptic curve with the order $|{\bf E}({\bf F}_q)|=l_1l_2$ where $l_1<l_2$ are two positive integers satisfying $\gcd(l_i, q)=1$, $i=1, 2$, and $\gcd(l_1, l_2)=1$. Then there exists an order $l_1$ subgroup ${\bf E}_1$ of ${\bf E}({\bf F}_q)$ and one coset ${\bf P}$ of ${\bf E}_1$, the elliptic curve code ${\bf C}({\bf P}, mP_0, {\bf E})$ is an MDS code for each $m$ satisfying $2\leq m \leq l_1-1$.}\\

{\bf Proof.} We have an elliptic curve ${\bf E}$ defined over ${\bf F}_q$ such that ${\bf E}({\bf F}_q)$ has a subgroup of the form ${\bf Z}/l_1l_2{\bf Z}$ from Theorem 2.2 and the condition $\gcd(l_1, q)=\gcd(l_2, q)=1$. From the condition $\gcd(l_1, l_2)=1$, the group ${\bf E}({\bf F}_q)$ is of the form ${\bf Z}/l_1{\bf Z} \oplus {\bf Z}/l_2{\bf Z}$. In this case, by setting ${\bf E}_1$ be the cyclic subgroup ${\bf Z}/l_1{\bf Z} \times {\bf 0}$ of the order $l_1$ and $b$ be the for ${\bf 0} \times x$, where $x$ is the generator of ${\bf Z}/l_2{\bf Z}$, the conclusion follows from Theorem 3.1.\\

{\bf Corollary 3.2.} {\em Let $n$ be a positive integer satisfying $6 \leq n \leq q^{1/4}$, $\gcd(n, q)=1$, and $m$ be any positive integer satisfying $2 \leq m \leq \frac{n}{2}$. Then there is an MDS elliptic curve code over ${\bf F}_q$ with the length $n$ and dimension $m$. This MDS code is not equivalent to the Reed-Solomon $[n, m, n-m+1]_q$ code when $m \geq 3$.}\\

{\bf Proof.} First of all, we can find an elliptic curve of the order $nl$ where $l$ is a positive integer satisfying $\gcd(n, l)=1$.  Then the conclusion follows from Theorem 3.1 and Theorem 2.4 immediately.\\

Since the general MDS conditions about twisted Reed-Solomon codes are restricted to subfields or subgroups as in \cite{BPR}, certainly many MDS codes constructed in Corollary 3.2 are not equivalent to these MDS twisted Reed-Solomon codes, or there is no known MDS twisted Reed-Solomon code with the corresponding length. Therefore many new non-Reed-Solomon type MDS codes from elliptic curves are constructed for consecutive lengths. We can observe the following example of MDS twisted Reed-Solomon codes as in \cite{BPR}. \\

Let $n$ and $k$ be two positive integers satisfying $n\ \leq q-1$ and $k \leq n-1$. Let $\alpha_1, \ldots, \alpha_n$ be $n$ distinct elements in the finite field ${\bf F}_q$ such that ${\bf \alpha}=\{\alpha_1, \ldots, \alpha_n\}$ is a $n$ element subset ${\bf F}_q$. Let $\eta$ be a nonzero element of ${\bf F}_q$. Set ${\bf g}_0=1+\eta x^k$, ${\bf g}_1=x$, \ldots, ${\bf g}_{k-1}=x^{k-1}$. Let ${\bf P}(\eta, k)$ be the linear span over ${\bf F}_q$ by ${\bf g}_0, \ldots, {\bf g}_{k-1}$. The linear $[n, k]_q$ code ${\bf C}_{{\bf \alpha}, \eta, k}$ is the evaluation code of these polynomials in ${\bf P}(\eta, k)$ at the above $n$ elements in the subset ${\bf \alpha}$. The dimension of the Schur square of ${\bf C}_{{\bf \alpha}, \eta, k}$ is at least $2k$. Thus this code is not equivalent to a Reed-Solomon code when $2k \leq n$. It is not hard to verify that if $\eta$ can not be represented as the product of any $k$ elements in ${\bf \alpha}$, that is, $$\eta \neq \prod_{1\leq j \leq k} \alpha_{i_j}, $$ for any $k$ distinct $\alpha_{i_1}, \ldots, \alpha_{i_k} \in {\bf \alpha}$, this code ${\bf C}_{{\bf \alpha}, \eta, k}$ is an MDS code. \\

The above condition about the set ${\bf \alpha}$ is strong, if we want to construct MDS twisted Reed-Solomon codes for large dimensions.  From the construction in Theorem 3.1, the MDS condition for elliptic curve codes is not so strong. Hence there are many MDS elliptic curve codes, which are not equivalent to known MDS twisted Reed-Solomon codes, or there is no known MDS twisted Reed-Solomon code with the corresponding length.\\

{\bf Corollary 3.3.} {\em Let $p$ be an odd prime number. Then there are MDS $[\lfloor \sqrt{p} \rfloor, k, \lfloor \sqrt{p} \rfloor-k+1]_p$ codes for $k=2, \ldots, \frac{\lfloor \sqrt{p} \rfloor}{2}$. There are MDS $[\lfloor \sqrt{p} \rfloor+1, k, \lfloor \sqrt{p} \rfloor-k+2]_p$ codes for $k=2, \ldots, \frac{\lfloor \sqrt{p} \rfloor}{2}$. These codes are not equivalent to Reed-Solomon codes when $k \geq 3$. }\\

{\bf Proof}. Set $n=\lfloor \sqrt{p} \rfloor$, it is clear that $$p-\sqrt{p} \leq n(n+1) \leq p+\sqrt{p},$$ $\gcd(n, p)=\gcd(n+1, p)=1$. Then there is an elliptic curve  ${\bf E}$ defined over ${\bf F}_p$ with $n(n+1)$ rational points from Theorem 2.1 and 2.2, with the group structure $|{\bf E}({\bf F}_p)={\bf Z}/n{\bf Z} \oplus {\bf Z}/(n+1){\bf Z}$. The conclusion follows from Theorem 3.1.\\

Now we can use supersingular elliptic curves to construct non-equivalent MDS codes. Let us recall some basic facts about supersingular elliptic curves. From \cite[page 152]{Silverman}, the elliptic curve ${\bf E}$ defined over ${\bf F}_p$ by $y^2=x^3+1$ is supersingular when $p \equiv 2$ $mod$ $3$, and the elliptic curve ${\bf E}$ defined over ${\bf F}_p$ by $y^2=x^3+x$ is supersingular when $p \equiv 3$ $mod$ $4$. For a supersinglular elliptic curve ${\bf E}$ defined over ${\bf E}_p$ it is known that $$|{\bf E}({\bf F}_{p^n})|=p^n+1,$$ if $n$ is an odd positive integer, or $$|{\bf E}({\bf F}_{p^n})|=(p^{\frac{n}{2}}-(-1)^{\frac{n}{2}})^2,$$ if $n$ is an even positive integer, see page 155 of \cite{Silverman}. We have the following result.\\

{\bf Corollary 3.4.} {\em Let $p$ be an odd prime satisfying $p \equiv 2$ $mod$ $3$ or $p \equiv 3$ $mod$ $4$. Let $N$ be a factor of $p^n+1$ satisfying $N < \sqrt{p^n+1}$, when $n$ is odd, and $N$ be a factor of $(p^{\frac{n}{2}}-(-1)^{\frac{n}{2}})^2$ satisfying $N < p^{\frac{n}{2}}-(-1)^{\frac{n}{2}}$ when $n$ is even. Then there is an MDS $[N, k, N-k+1]_{p^n}$ linear code. This MDS code is not equivalent to the Reed-Solomon code when $k \leq \frac{N}{2}-1$.}\\

{\bf Proof.} There is no factor $p^l$ in $|{\bf E}({\bf F}_{p^n})|$, then we can take $a(l)=0$ for any prime factor $l$ of $|{\bf E}({\bf F}_{p^n})|$ in Theorem 2.2. The conclusion follows from Theorem 3.1 1) directly.\\

Some non-Reed-Solomon MDS codes were introduced in \cite{RL1989}, it is not hard to verify that the minimum Hamming distances of the Schur square of Roth-Lemple MDS codes are very small. Hence many of our constructed MDS elliptic curve codes are not equivalent to Roth-Lempel MDS codes.\\

Moreover from Theorem 2.3 and 2.4,  we can construct some MDS elliptic curve codes such that their Schur squares are MDS codes or not MDS codes.  Therefore some non-equivalent MDS elliptic curve codes from one elliptic curve or two different elliptic curves (with different elliptic curve group structures) of the same length and the same dimension can be constructed.\\

\section{Non-Reed-Solomon type MDS codes from higher genus $g \geq 2$ curves}

In this section, we prove that MDS codes from a genus $g>g'$ curve is not equivalent to MDS codes from a genus $g'$ curve, when $m\geq 4g$. Then MDS codes from genus $g\geq 2$ curves are essentially new if the degree of ${\bf G}$ is bigger than or equal to $4g$.  It is proved that at least short length new non-Reed-Solomon type MDS codes from higher genus $g \geq 2$ curves can be constructed.\\

The following result is a direct generalization of Theorem 2.4.\\

{\bf Theorem 4.1.} {\em Let ${\bf X}$ be a genus $g$ curve defined over ${\bf F}_q$, $P_0, P_1, \ldots, P_n$ are $n+1$ rational point of ${\bf X}$, $m$ is a positive integer satisfying $4g<m<n$ and $2m<n$. Then the dimension of the Schur square of the dimension $k$ one point function code ${\bf C}(P_1, \ldots, P_n, mP_0, {\bf X})$ is exactly $2k+g-1$. In particular, MDS AG codes from a genus $g_1$ curves are not equivalent to MDS AG codes from a genus $g_2$ curve if $g_1>g_2$.}\\

{\bf Proof.} We recall the well-known Weierstrass gap theorem, see Chapter 6 of \cite{HKT}, except $g$ positive integers $1=\alpha_1<\alpha_2< \cdots <\alpha_g \leq 2g-1$, there is a rational function $f \in {\bf L}(mP_0)$ such that the pole part of the divisor $(f)$ associated with the function $f$ is exactly $m'P_0$, where $m' \neq \alpha_i$, $i=1, \ldots, g$. Therefore for each positive integer $m' \geq 2g$ we can find a rational function $f \in {\bf L}(mP_0)$ such that the pole part of $f$ is exactly $m'P_0$. Then the Schur square of ${\bf C}(P_1, \ldots, P_n, mP_0, {\bf X})$ is exactly ${\bf C}(P_1, \ldots, P_n, 2mP_0, {\bf X})$ with the dimension $2m-g+1=2(m-g+1)+g-1=2k+g-1$.\\

Now we observe the MDS condition for AG codes from genus $g$ curves. In the case $g=1$, this is just Theorem 2.3 1).\\

{\bf Theorem 4.2.} {\em  Let ${\bf X}$ be a genus $g$ curve defined over ${\bf F}_q$, $P_0, P_1, \ldots, P_n$ are $n+1$ rational point of ${\bf X}$, $m$ is a positive integer satisfying $2g-1<m<n$. Then the one point function code ${\bf C}(P_1, \ldots, P_n, mP_0, {\bf X})$ is MDS if and only if the following MDS condition holds.\\

{\bf MDS condition for AG codes}: For every $m-g+1=k$ different points $P_{i_1}, \ldots, P_{i_k}$ among $P_1, \ldots, P_n$, there is no degree $g-1$ rational effective divisor ${\bf G}'$ such that the divisor $P_{i_1}+\cdots+P_{i_{m-g+1}}+{\bf G}'$ is linear equivalent to the divisor $mP_0$.}\\

{\bf Proof.} We only need to prove that there is no weight $w$ codeword, where $n-m \leq w \leq n-m+g-1$. This is obvious since there is no weight $n-m+i$ codeword in this function code for $i=0, \ldots, g-1$, from the above MDS condition.\\

The existence of such $n$ rational points $P_1, \ldots, P_n$ satisfying the above MDS condition  can be proved by a counting argument, at least for small $n \geq 5g$. We can consider the above MDS condition for the projective imbedding $\Phi$ of the curve ${\bf X}$ in ${\bf P}^{m-g}$ defined by the divisor $mP_0$. Let $\Phi({\bf X})$ be the image of this curve ${\bf X}$ in ${\bf P}^{m-g}$. Then the above MDS condition is equivalent to the following condition, also see \cite{Walker}.\\

{\bf MDS condition for embedding}: To find $n$ rational points $P_1, \ldots, P_n$ in $\Phi({\bf X})$, such that there is no $m-g+1$ different points $P_{i_1}, \ldots, P_{i_{m-g+1}}$ among them, which are in a hyperplane of ${\bf P}^{m-g}$.\\

Therefore the following result is direct from a simple counting argument. The existence of short length MDS codes from higher genus curves is proved.\\

{\bf Theorem 4.3.} {\em Let ${\bf X}$ be a genus $g \geq 2$ curve defined over ${\bf F}_q$, and $m$ be a positive integer satisfying $m\geq 4g$. If $\Phi({\bf X}) \subset {\bf P}^{m-g}$ defined by the linear system $mP_0$ is a non-singular curve with $N$ rational points which are images of rational points of ${\bf X}$. If $n$ is a positive integer satisfying $m<n$ and $m \cdot \displaystyle{n \choose m-g} <N$, then there exists a length $n$ and dimension $m-g+1$ MDS code from ${\bf X}$. These MDS codes are not equivalent to Reed-Solomon codes.}\\

{\bf Proof.} For each different $m-g$ rational points in general position amonge chosen evaluation points, we determine a hyperplane in ${\bf P}^{m-g}$. This hyperplane interests $\Phi({\bf X})$ at $m$ points. Then the conclusion follows directly.\\

In particular, when $m=O(g)$ is fixed then new non-Reed-Solomon MDS codes of lengths $n \leq O(q^{\frac{1}{m-g}})$ over ${\bf F}_q$ of the dimension $m-g+1$ can be constructed from Theorem 4.3, when $q$ tends to the infinity. \\

It would be interesting to construct explicit longer MDS AG codes from the Hermitian curve, recent constructed maximal curves in\cite{PM}. This could give explicit new non-Reed-Solomon type MDS codes from higher genus curves, which are longer than MDS codes from Theorem 4.3. It is also easy to generalize Theorem 4.1 to two point AG codes, see \cite{BBDNR,LV}, and to construct more new longer non-Reed-Solomon type MDS codes from higher genus curves.\\

\section{Self-dual MDS codes from elliptic curves over finite fields ${\bf F}_{2^s}$.}

In this section we restrict to the elliptic curves ${\bf E}$ defined over a finite field ${\bf F}_{2^s}$ of the characteristic $2$. The set ${\bf P}=\{P_1, \ldots, P_n\}$ is one coset of a subgroup ${\bf E}_1 \subset {\bf E}({\bf F}_{2^s})$. We show that there are many self-dual MDS elliptic curve codes from elliptic curves ${\bf E}$ defined over ${\bf F}_{2^s}$. We always are in the case a) of Theorem 2.1. That is, $\beta$ is an odd positive integer satisfying that $\gcd(2-\beta, 2^s-1)>1$ has an odd divisor. Then the group order $|{\bf E}({\bf F}_{2^s})|=2^s+1-\beta$ is an even number, and $2^s+1-\beta=2^s-1+2-\beta$, $\gcd(|{\bf E}({\bf F}_{2^s})|, 2^s-1)>1$ has an odd divisor. Actually since $\beta$ can be any odd number in the case a) of Theorem 2.1, when $q=2^s$, if $s$ is a composite number $s=s_1s_2$ it is obvious we can take $2-\beta \equiv 0$ $mod$ $(2^{s_1}-1)$. Hence there are many such elliptic curves with the desired group orders.\\

Therefore we need the following conditions to construct self-dual MDS elliptic curve codes.\\
1) $q=2^{s_1s_2}$, $2-\beta \equiv 0$ $mod$ $2^{s_1}-1$;\\
2) $\beta\equiv 1$ $mod$ $8$;\\
3) An elliptic curve defined over ${\bf F}_{2^{s_1s_2}}$ with $2^{s_1s_2}+1-\beta$ rational points;\\

Since $2^{s_1s_2}+1-\beta \equiv 0$ $mod$ $8$, the exponent $h_2$ of $2$ in the prime factor decomposition of $$|{\bf E}({\bf F}_{2^{s_1s_2}})|=2^{s_1s_2}+1-\beta,$$ is at least $h_2 \geq 3$. \\

4) The above elliptic curve is of the group structure ${\bf Z}/2^{h_2}{\bf Z} \oplus{\bf Z}/L{\bf Z} $, where $h_2 \geq 3$ and $L$ is an odd positive integer.\\

From Theorem 2.1 and 2.2 there are many such an elliptic curve. Then the group order of the elliptic curve is $$|{\bf E}({\bf F}_{2^{s_1s_2}})|=2^{h_2}L,$$ where $h_2 \geq 3$ and $L$ is an odd positive integer.\\

{\bf Theorem 5.1.} {\em Let $q=2^{s_1s_2}$ as above and $L'$ be an odd divisor of $L$, $n=2^tL'$ where $t \leq h_2-1$. There exists a self-dual $[n, \frac{n}{2}, \frac{n}{2}+1]_{2^{s_1s_2}}$ MDS code equivalent to an elliptic curve code from ${\bf E}$,  which is not equivalent to the Reed-Solomon code.}\\

{\bf Proof.} From Theorem 2.2, we can find elliptic curve such that for any odd divisor $L'|L$, there is an order $L'$ subgroup ${\bf E}_2 \subset {\bf E}({\bf F}_{2^{s_1s_2}})$. Therefore we have an order $2^t L'$ subgroup ${\bf E}_1$ of the form $(i2^{h_2-t} \theta) \times {\bf E}_2$, where $\theta$ is the generator of the cyclic subgroup ${\bf Z}/2^{h_2}{\bf Z} \subset {\bf E}({\bf F}_{2^{s_1s_2}})$, and $i=0, 1, \ldots, 2^t-1$.\\

Then set $b=2^{h_2-1-t} \theta \times {\bf 0} \subset {\bf E}({\bf F}_{2^{s_1s_2}})$. The coset ${\bf P}=b+{\bf E}_1$ has $2^tL'$ elements. The sum of arbitrary $\frac{n}{2}$ different elements in ${\bf P}$ is of the form $2^{t-1}L' b+P$, where $P$ is an element in the subgroup ${\bf E}_1$. It is clear that $2^{t-1}L' b+P$ is of the form $$2^{h_2-2}L'\theta \times {\bf 0}+P.$$ This is not zero. Therefore the elliptic curve code ${\bf C}({\bf P}, 2^tL'P_0, {\bf E})$ is an MDS code from Theorem 2.3. Here $P_0$ is the zero element of the group ${\bf E}({\bf F}_{2^{s_1s_2}})$.\\

The sum of all elements in the coset ${\bf E}_1$ is of the form $-2^{h_2-1} L'\theta \times {\bf 0}$, since the sum of all elements in an order $L'$ subgroup, $L'$ odd, is zero, and the sum of all elements in ${\bf Z}/2^u{\bf Z}$ is $-2^{u-1}$. Then the sum of all elements in ${\bf P}$ is $2^{h_2-1}L'\theta  \times {\bf 0}-2^{h_2-1} L'\theta \times {\bf 0}$ is zero element $P_0$. Therefore from Theorem 2.3. 2), the dual code of ${\bf C}({\bf P}, 2^{t-1} L' P_0, {\bf E})$ is equivalent to a linear code ${\bf  C}({\bf P}, 2^{t-1}L'P_0, {\bf E})$. Suppose that the dual code is of the form $${\bf v} \cdot {\bf C}({\bf P}, 2^{t-1}L'P_0, {\bf E})=\{(v_1c_1, \ldots, v_n c_n): (c_1, c_2, \ldots, c_n) \in {\bf C}({\bf P}, 2^{t-1}L'P_0, {\bf E}) \},$$ where ${\bf v}=(v_1, \ldots, v_n) \in {\bf F}_{2^{s_1s_2}}^n$ is a Hamming weight $n$ vector.\\

Since this field is of characteristic $2$, each element $v_i$ is a square, set $v_i=v_i'^2$, $i=1, \ldots, n$. Set ${\bf v'}=(v_1', \ldots, v_n')$, the equivalent code ${\bf v'} \cdot {\bf C}({\bf P}, 2^{t-1}L'P_0, {\bf E})$ is a self-dual code. Actually $({\bf v'} \cdot {\bf C}({\bf P}, 2^{t-1} L' P_0, {\bf E}))^{\perp}=\frac{1}{{\bf v'}} \cdot {\bf C}({\bf P}, 2^{t-1}L'P_0, {\bf E})^{\perp}={\bf v'} \cdot {\bf C}({\bf P}, 2^{t-1}L'P_0, {\bf E})$. The dimension of the Schur square of this self-dual code is exactly $n$. It is not equivalent to a Reed-Solomon $[n, \frac{n}{2}, \frac{n}{2}+1]_{2^{s_1s_2}}$ code, the dimension of its Schur square is $n-1$. The conclusion is proved.\\

Notice that $L'$ can be any odd divisor of the group order $|{\bf E}({\bf F}_{2^{s_1s_2}})|$, there are indeed many self-dual MDS codes which are equivalent to elliptic curve codes. Actually self-dual MDS elliptic curve codes over ${\bf F}_{2^s}$ of the length $4L$, where $L$ is any odd positive number in the range $[1, \frac{2^s+1+\lfloor 2\sqrt{2^s} \rfloor}{8}]$, can be constructed. Hence there are many new self-dual MDS elliptic curve codes over the finite field ${\bf F}_{2^s}$, which are not equivalent to self-dual Reed-Solomon codes or self-dual twisted Reed-Solomon codes.\\

From the result in \cite{HChen2}, it is easy to construct equivalent LCD MDS codes from self-dual MDS elliptic curve codes.\\

From Theorem 5.1 and the CSS construction of EAQEC codes in \cite{Brun}, the following results follows immediately.\\

{\bf Corollary 5.1.} {\em Let $q=2^{s_1s_2}$  be an even prime power and $|{\bf E}({\bf F}_q)|=2^{h_2} \cdot L$ as in above, $n$ be a positive integer of the form $2^tL'$ where $L'$ is an odd divisor of the group order ${\bf E}({\bf F}_{2^{s_1s_2}})$ and $t \leq h_2-1$, and $k$ be a positive integer satisfying $\frac{n}{2} \leq k \leq n-1$, and $h$ be a nonnegative integer satisfying $0 \leq h \leq \frac{n}{2}$, there exists an MDS EAQEC $[[n, k-h, n-k+1, n-k-h]]_{2^{s_1s_2}}$ code.}\\

{\bf Proof.} From Theorem 5.1 we have an equivalent MDS self-dual elliptic curve $[n, \frac{n}{2}, \frac{n}{2}+1]_{2^{s_1s_2}}$ code. From the result in \cite{HChen2} we have a linear MDS $[n, \frac{n}{2}, \frac{n}{2}+1]_{2^{s_1s_2}}$ code with the $h$-dimension hull, where $0 \leq h \leq \frac{n}{2}$. Then the conclusion follows from the CSS construction of EAQEC codes immediately.\\

Notice that the MDS elliptic curve codes in Theorem 5.1 are not equivalent to Reed-Solomon code, then these MDS EAQEC codes in Corollary 5.1 are new MDS EAQEC codes, comparing to previous constructed MDS EAQEC codes from generalized Reed-Solomon codes.\\

\section{Conclusion and unsolved problems}

In this paper, we proved that there are many MDS AG codes from arbitrary genus algebraic curves, and self-dual MDS AG codes from elliptic curves. These MDS codes are not equivalent to Reed-Solomon codes, not equivalent to known MDS twisted Reed-Solomon codes and not equivalent to Roth-Lempel MDS codes. This showed that MDS AG codes can be obtained from carefully chosen evaluation points in an arbitrary genus curve. The following two problems about longer MDS AG codes seem interesting.\\

1) What is the range of lengths such that there are non-Reed-Solomon type MDS AG codes. Ranges of lengths of MDS elliptic curve codes in Corollary 3.3 and 3.4 are obviously not optimal. Are there non-Reed-Solomon type  MDS codes from elliptic curve over prime field ${\bf F}_p$ with any given length $n \leq \frac{p}{3}$?\\

2) Can longer explicit MDS $[n, k, n-k+1]_q$ codes be constructed as one point AG codes from genus $g\geq 2$ curves for each genus $g \geq 2$?  We proved that there are very short length MDS codes from higher genus curves. It is interesting to construct longer MDS AG codes from maximal curves, with explicit evaluation points. \\

\end{document}